\documentclass[11pt]{article}
\pdfoutput=1
\usepackage[a4paper]{geometry}
\usepackage{color}
\usepackage{float}
\usepackage[utf8]{inputenc}
\usepackage{a4wide}
\usepackage{amsmath,amssymb,amstext,amsthm,amssymb}
\usepackage{amsfonts}
\usepackage{framed}
\usepackage{graphicx}
\usepackage{bbold}
\usepackage{bbm}
\usepackage{slashed} 
\usepackage{tensor} 
\usepackage{braket} 
\usepackage{amstext}
\usepackage{array}
\usepackage{graphicx}
\usepackage{setspace}
\usepackage{hyperref}
\usepackage{overpic}
\usepackage{bbm}
\usepackage{cite}
\usepackage{lipsum}
\usepackage{tikz}

\newcommand{\be}{\begin{equation}}
\newcommand{\ee}{\end{equation}}
\newcommand{\bea}{\begin{eqnarray}}
\newcommand{\eea}{\end{eqnarray}}

\begin{document}

\title{WGC for Axions}

\pagenumbering{gobble}
\begin{center}
{\huge \textbf {What is the} \\ 
\textbf{Magnetic Weak Gravity Conjecture} \\ 
\vspace{0.35cm}
\textbf{for Axions?}}\\
\vspace{1.5cm}
{\large Arthur Hebecker$^1$, Philipp Henkenjohann$^1$ and Lukas T. Witkowski$^2$}\\
\vspace{0.5cm}
\textit{$^1$ Institute for Theoretical Physics, University of Heidelberg, \\
Philosophenweg 19, 69120 Heidelberg, Germany\\
\vspace{0.3cm}
$^2$ APC, Universit\'e Paris 7, CNRS/IN2P3, CEA/IRFU, Obs.~de Paris,
Sorbonne Paris Cit\'e, B\^atiment Condorcet, F-75205, Paris Cedex 13, France
(UMR du CNRS 7164)}\\
\vspace{1cm}
\textbf{Abstract}\\
\end{center}
\vspace{0.5cm}
The electric Weak Gravity Conjecture demands that axions with large decay constant $f$ couple to light instantons. The resulting large instantonic corrections pose problems for natural inflation. We explore an alternative argument based on the magnetic Weak Gravity Conjecture for axions, which we try to make more precise. Roughly speaking, it demands that the minimally charged string coupled to the dual 2-form-field exists in the effective theory. Most naively, such large-$f$ strings curve space too much to exist as static solutions, thus ruling out large-$f$ axions. More conservatively, one might allow non-static string solutions to play the role of the required charged objects. In this case, topological inflation would save the superplanckian axion. Furthermore, a large-$f$ axion may appear in the low-energy effective theory based on two subplanckian axions in the UV. The resulting effective string is a composite object built from several elementary strings and domain walls. It may or may not satisfy the magnetic Weak Gravity Conjecture depending on how strictly the latter is interpreted and on the cosmological dynamics of this composite object, which remain to be fully understood. Finally, we recall that large-field brane inflation is naively possible in the codimension-one case. We show how string-theoretic back-reaction closes this apparent loophole of large-$f$ (non-periodic) pseudo-axions.

\vspace*{10ex}
\noindent January 23, 2017
\newpage
\pagenumbering{arabic}
\section{Introduction}
The Weak Gravity Conjecture (WGC) \cite{0601001} provides a condition for identifying low energy effective theories which do not permit a UV-completion and should hence be assigned to the `swampland' \cite{0509212, 0605264}. If true, it is a powerful tool with applications to both particle physics and cosmology. In its original form, the WGC is a statement regarding the electric and magnetic particle content of a $U(1)$ gauge theory coupled to gravity. It can be extended to encompass theories with a gauge group consisting of multiple $U(1)$ factors \cite{1402.2287} and with charged $p$-branes rather than just particles \cite{0601001, 1503.04783, 1509.06374}. 

A particularly interesting extension of the WGC is that to $(-1)$-branes. The resulting WGC for axions and instantons potentially constrains large-field inflation in the field space of one or multiple axions \cite{1409.5793, 1412.3457, 1503.00795, 1503.03886, 1503.04783, 1503.07853, 1503.07912, 1504.00659, 1504.03566, 1506.03447, 1509.07049} (see also \cite{0303252} for constraints motivated by string theory). In addition, models of axion monodromy inflation (and the related Cosmological Relaxation mechanism) can be constrained by the WGC for domain walls \cite{1503.04783,1512.00025, 1512.03768}. For further recent work on the WGC see also \cite{1509.01647, 1510.07911, 1610.01533, 1610.04564, 1611.01395, 1611.08953}.

In fact, the WGC can be given in its magnetic and electric form, which are a priori two independent statements. The electric WGC constrains the electrically charged spectrum of the theory. It is motivated by requiring the absence of stable black holes \cite{0601001} (see \cite{161106270} for progress in making this logic more rigorous). The magnetic WGC arises from requiring the minimally charged magnetic object not to be a black hole or black brane. It can be phrased as a an upper bound for the UV cutoff of the theory. To be specific, for a $(p+1)$-form gauge theory in $d$ dimensions with electrically charged $p$-branes the magnetic WGC requires 
\begin{align}
\label{eq:generalcutoff} \Lambda \lesssim {\left(g_\text{e} M_\text{P}^{\frac{d}{2}-1} \right)}^{\frac{1}{p+1}} \, ,
\end{align}
where $g_\text{e}$ is the coupling constant of the electric theory. For the case of a theory with particles ($p=0$) in $4$ dimensions this reduces to the well-known statement $\Lambda \lesssim g_\text{e} M_\text{P}$. 

Most constraints on axion inflation arise from the electric WGC for axions. It relates the axion periods to the size of instantonic terms in the axion potential, thus directly constraining the inflaton potential. In some studies also the magnetic WGC has been invoked to constrain inflation. For example, 5-dimensional models of extranatural inflation are constrained by the magnetic WGC for particles \cite{1412.3457} while the magnetic WGC for domain walls is relevant for axion monodromy \cite{1512.03768}. However, it is not the WGC for axions which is employed in these cases. In this paper, we want to focus on precisely this possibility, i.e.~the role of the magnetic WGC for axions.

A naive approach for arriving at a magnetic WGC for axions is to consider \eqref{eq:generalcutoff}. Note that for the case of an axion ($p=-1$) the exponent in \eqref{eq:generalcutoff} diverges. In 4 dimensions the electric coupling is given by $g_\text{e} = 1/f$ where $f$ is the axion decay constant determining the axion period.  The r.h.~side of \eqref{eq:generalcutoff} involves $(M_\text{P}/f)$ raised to a diverging power. Clearly, for $f<M_\text{P}$ this gives no constraint on the cutoff. By contrast, for $f>M_\text{P}$ the inequality implies $\Lambda\to 0$, i.e.~the theory simply does not exist. As $f > M_\text{P}$ is necessary for large-field inflation, this would imply that the magnetic WGC censors large-field axion inflation. 

In what follows, we want to go beyond this simple estimate and develop a more rigorous argument based of the magnetic WGC. As the magnetic WGC is concerned with the existence and the properties of magnetically charged objects, we will study these objects in detail. For an electric theory of instantons coupled to an axion, the corresponding magnetic object is a string coupled to a 2-form field. Thus, in this work, we will study (cosmic) string solutions in a theory with $f > M_\text{P}$ explicitly. We take the following statement as the preliminary definition of the magnetic WGC for axions: For the magnetic WGC to be satisfied we require the minimally charged (cosmic) string to exist as a field-theoretic object. The task is hence to examine whether string solutions for $f > M_\text{P}$ suffer from any pathologies. Our most promising candidates will be topological
inflation and and a composite string \`a la \cite{1606.05552,1608.06951}.

\subsubsection*{Summary of results}
We begin by analysing the static spacetime solution of the exterior of an axionic string \cite{Cohen:1988sg} (see \cite{0510033, 1011.5120, 9502069} for discussions related to such strings from a string theory perspective). There are two problems which put the existence of this object in doubt for $f > M_\text{P}$.
\begin{itemize}
\item There is a singularity at a finite distance from the string core. This singularity even persists for $f < M_\text{P}$, but would then be exponentially far away.
\item For $f > M_\text{P}$ the deficit angle around the string is always negative. This makes it impossible to attach such a spacetime to that of a proper UV-completion of the string core since we expect this to be locally flat at the string's center. In contrast, for $f < M_\text{P}$ there would have been a finite region with a positive deficit angle as expected in the vicinity of a string.
\end{itemize}
Hence, the \emph{static} string solution \cite{Cohen:1988sg} does not seem to exist for $f > M_\text{P}$ and thus does not provide us with the magnetic object required for satisfying the WGC. 

The problem of the singularity at a finite distance can be resolved by considering a string solution with a dynamical spacetime \cite{9606002, 0208037}. Interestingly, this dynamical solution exists for $f > M_\text{P}$ up to a maximal value $f_{\textrm{max}}$ which lies in the range $6 \leq f_{\textrm{max}}^2 /M_\text{P}^2 \leq 12$. However, it remains questionable whether the dynamical solution can be interpreted as a string for $f>M_\text{P}$. Instead, for  $f>M_\text{P}$ the Hubble length becomes comparable to the size of the string core leading to an expansion of the defect in all directions. This is known as topological inflation \cite{9402031,9402085}. It then remains to be checked whether this solution can play the role of a bona fide string for an observer outside the inflating core.

However, accepting topological inflation as a UV-completion leads to further puzzles. Topological inflation can also arise in 1-form theories where the inflating defect is a magnetic monopole. For example, consider the theory giving rise to such monopoles used by Linde in \cite{9402031}. This is an SU(2) YM-theory which is broken down to U(1) by an adjoint scalar. Let $g\ll1$ be the gauge coupling, $\lambda$ the scalar field self-coupling, and $v$ the symmetry breaking vev. The size of the magnetic monopole is given by the maximum of the two length scales $(gv)^{-1}$ and $(\sqrt{\lambda}v)^{-1}$ where $gv$ is the gauge boson mass and $\sqrt{\lambda}v$ is the scalar mass. The corresponding cutoff scale $\Lambda$ of the monopole is $\text{min}\{gv,\sqrt{\lambda}v\}$. Let us try to make $\Lambda$ as large as possible while keeping the theory under control, i.e.~let us make sure that energy densities remain subplanckian. We must therefore ensure that $\lambda v^4\le M_\text{P}^4$. This optimization problem is solved by $\sqrt{\lambda}=g$ and $v=M_\text{P}/\sqrt{g}$. The resulting cutoff $\Lambda=\sqrt{g}M_\text{P}$ is subplanckian but exceeds the WGC bound $gM_\text{P}$. Since $v>M_\text{P}$, the corresponding solution will inflate. If we accept this as the UV-completion of the minimally charged monopole required by the WGC, the argument for the low cutoff, $\Lambda\sim gM_\text{P}$, is lost. Indeed, the cutoff explicitly found above is higher. One is then left with the unsatisfactory situation that topological inflation ensures that the magnetic WGC remains satisfied in axion theories, while for 1-form theories it would allow for an explicit violation. This can be avoided by slightly strengthening the requirement underlying the magnetic WGC. Note that besides the inflating monopole solution a minimally charged BH-monopole still exists in our setting. If the correct formulation of the magnetic WGC is that no minimal magnetic monopole should be a BH, the above situation remains forbidden. One might then be tempted to conclude that topological inflation resides in the swampland.

Thus, having studied both static and dynamical string solutions, our analysis leaves us with the following results:
\begin{itemize}
\item  If topologically inflating spacetimes are acceptable as the magnetic objects required by the WGC, then $f > M_\text{P}$ cannot be ruled out in the present approach. 
\item  By contrast, if topologically inflating spacetimes are not accepted as magnetically charged objects of axion models, we have good reason to believe that $f>M_\text{P}$ is forbidden. Such a viewpoint can be argued from the fact that the relevant solutions are non-static or from the presence of a horizon. The conclusion concerning large $f$ would then be as negative as in the naive approach mentioned in the beginning.
\end{itemize}

Last, we turn to axion theories with $f_{\textrm{eff}} > M_\text{P}$ which admit a UV completion in terms of a theory of two axions with $f_{1,2} < M_\text{P}$ \cite{0409138,0912.1341,1407.2562}. As suggested in \cite{1503.07912} using a stringy example, a particularly promising way to realize the required winding trajectory is through 3-form gauging \cite{0507215} (for applications of this approach in the context of axion monodromy inflation see \cite{0811.1989,1101.0026}).\footnote{This corresponds to the familiar St\"uckelberg mechanism in the 1-form case and can be used to obtain a small effective gauge coupling \cite{1608.06951}.}As $f_{1,2} < M_\text{P}$, the corresponding magnetic objects exist and avoid the problems mentioned above. Following \cite{1606.05552,1608.06951}, one can then identify a string solution as bound state of the strings of the original two axions connected by domain walls. This object is then checked explicitly for possible pathologies.

One observation is that the tension of this effective string becomes superplanckian if $f > M_\text{P}$. The tension can be estimated as $T_\text{eff} \sim f_\text{eff}^2$ and hence one leaves the regime of weak gravity back-reaction once $f_\text{eff}$ is sufficiently large. A modestly super-Planckian $f_\text{eff}$ may be possible, but this will depend on the precise numerical factors. Our finding is consistent with the previous results for explicit string solutions. Static string solutions for parametrically large $f$ do not seem to exist. Dynamical solutions are possible, but we have to leave their detailed study in this particular case to future work. While they may be similar to topological inflation, they could equally produce a singularity or exhibit a completely different, unexpected behaviour.

Very recently the paper \cite{1701.05572} appeared which has some overlap with and is in part complementary to the present work.

\section{Naive estimates}

The magnetic WGC follows from the requirement that the minimally charged magnetic object of a U(1) gauge theory is not a black hole. The mass $M$ of a monopole in 4 dimensions can be estimated by its field energy, leading to $M\sim g_\text{m}^2\Lambda$, where $g_\text{m}$ denotes the magnetic coupling constant and $\Lambda$ is a cutoff that determines the radius of the core. The electric coupling constant is related to $g_\text{m}$ by $g_\text{e}\sim g_\text{m}^{-1}$. If the core radius is larger than the Schwarzschild radius, the monopole is not a black hole. The corresponding formula is $\Lambda^{-1}\ge M/M_\text{P}^2\sim g_\text{e}^{-2}\Lambda/M_\text{P}^2$, implying $\Lambda \lesssim g_\text{e}M_\text{P}$.

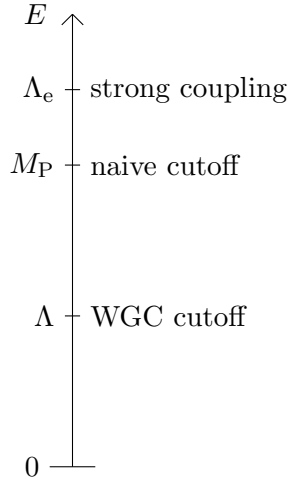
\begin{figure}
\begin{center} 
\begin{tikzpicture}
\draw (-0.2,6) node[anchor=east] {$E$};
\draw (0,0)--(0,6);
\draw (0,6)--+(-45:0.2);
\draw (0,6)--+(-135:0.2); 
\draw (-0.3,0) node[anchor=east] {0} --(0.3,0);
\draw (-0.1,4) node[anchor=east] {$M_\text{P}$} --(0.1,4) node[anchor=west] {naive cutoff};
\draw (-0.1,5) node[anchor=east]{$\Lambda_\text{e}$} --(0.1,5) node[anchor=west] {strong coupling};
\draw (-0.1,2) node[anchor=east] {$\Lambda$} -- (0.1,2) node[anchor=west] {WGC cutoff};
\end{tikzpicture}
\caption{Cutoff scales of a weakly coupled gauge theory in the presence of gravity.}
\label{scales} 
\end{center}
\end{figure}

This easily generalizes to an `electric' $(p+1)$-form gauge theory with `magnetic' $(d-p-4)$-branes in $d$ dimensions. The electric coupling constant $g_\text{e}$ of this system has mass dimension $p+2-d/2$ and the magnetic coupling is $g_\text{m}\sim g_\text{e}^{-1}$. The field energy of the magnetic $(d-p-4)$-brane is proportional to $g_\text{m}^2$ and therefore, analogously to the monopole case, we can estimate the tension by

\be
T\sim g_\text{m}^2\Lambda^{p+1},
\ee
using dimensional analysis. On the other hand, the tension of a black brane is proportional to the inverse coupling constant $\kappa_d^{-2}=M_\text{P}^{d-2}$ of gravity. Hence

\be
T_\text{BH}\sim M_\text{P}^{d-2}R_\text{S}^{p+1},
\ee
where $R_\text{S}$ is the Schwarzschild radius. According to the magnetic WGC we need to impose $\Lambda^{-1}\ge R_\text{S}$. Expressing $R_\text{S}$ through $T_\text{BH}$ and using $T_\text{BH}\sim T$ gives

\be
\Lambda\lesssim (g_\text{e}^2M_\text{P}^{d-2})^{\frac{1}{2(p+1)}}.
\ee
We find it particularly intuitive to rewrite this in terms of the `characteristic energy scale' or `strong coupling scale' $\Lambda_\text{e}$ of the electric gauge theory, defined by $g_\text{e}^2=\Lambda_\text{e}^{2(p+2)-d}$. One finds

\be
\frac{\Lambda}{M_\text{P}}\lesssim \left(\frac{M_\text{p}}{\Lambda_\text{e}}\right)^{\frac{d-2(p+2)}{2(p+1)}}. \label{mwgcalt}
\ee
If $2(p+2)-d<0$, the gauge theory is IR-free and $\Lambda_\text{e}$ is its intrinsic UV cutoff. For small enough coupling, the energy scale $\Lambda_\text{e}$ exceeds $M_\text{P}$ and naively $M_\text{P}$ would now be the cutoff of the theory. However, if $p+1>0$, the magnetic WGC (\ref{mwgcalt}) predicts a cutoff that is smaller than the Planck mass. This situation is sketched in Fig.~\ref{scales}. The condition $p+1>0$ can be rewritten as $d_\text{co}>2$, where $d_\text{co}\equiv p+3$ is the co-dimension of the magnetic brane.

If $d_\text{co}=2$ the picture in Fig.~\ref{scales} applies but in an extreme case: The exponent in (\ref{mwgcalt}) diverges and, in the weakly coupled case $\Lambda_\text{e}>M_\text{P}$, the cutoff $\Lambda$ vanishes. Such a situation might be interpreted by saying that the theory does not exist, in other words, the weakly coupled case is forbidden. This occurs for example for the string ($p=1$) in 4 dimensions where the magnetic coupling constant is given by the axion decay constant $f=g_\text{m}$. Weak coupling corresponds to $f>M_\text{P}$ in this case, which should hence be impossible to realize. In the following we want to check this statement by trying to explicitly construct a string with $f>M_\text{P}$ in 4 dimensions. 

Before doing so, let us perform a very rough calculation in order to gain some intuition for what one can expect from a more detailed analysis. The field energy density of the string is $\rho\sim f^2/r^2$. This gives

\be
T\sim\int_0^{2\pi}\text{d}\varphi\int_{\Lambda_\text{UV}^{-1}}^{\Lambda_\text{IR}^{-1}}r\text{d}r\rho\sim f^2\text{ln}(\frac{\Lambda_\text{UV}}{\Lambda_\text{IR}}), \label{flatt}
\ee
where we had to introduce two cutoff scales $\Lambda_\text{UV}$ and $\Lambda_\text{IR}$ in order to render the contribution to the string tension finite. Let us now discuss the inclusion of gravity. The static vacuum solution of Einstein's equations with cylindrical symmetry has a deficit angle $\Delta\phi$ in the plane perpendicular to the symmetry axis, i.e.~it describes a cone:
\be
\text{d}s^2=-\text{d}t^2+\text{d}z^2+\text{d}r^2+r^2(1-\frac{\Delta\phi}{2\pi})^2\text{d}\theta^2. \label{conest}
\ee
This is the exterior spacetime of a string without charge \cite{Hiscock:1985uc} and

\be
\Delta\phi=\frac{T}{M_\text{P}^2}. \label{defangle}
\ee
For $T\gtrsim M_\text{P}^2$, i.e.~$\Delta\phi\gtrsim2\pi$, this spacetime breaks down (see \cite{Gott:1984ef, Laguna:1989rx, Linet:1990fk} and in particular \cite{Ortiz:1990tn} for a good review). Therefore, since (\ref{flatt}) indicates that $T\gtrsim f^2$, one can expect that corresponding axionic string spacetimes do not exist.

In fact, one can say more: Imagine that the field of the charged string is switched off at a distance $R$ from the string. The corresponding total tension of this configuration is estimated by (\ref{flatt}), where the upper integration limit is now given by $R$. At distances $r>R$ the vacuum solution (\ref{conest}) describes this spacetime, with deficit angle given by (\ref{defangle}). Now, repeating this argument for another radius $R'>R$ one immediately sees that the corresponding deficit angle is larger than that for $R$. We therefore see that the deficit angle of the spacetime of a charged string is not constant but grows with the distance to the string. Thus, one expects a spacetime which is locally conical buy eventually breaks down when $\Delta\phi>2\pi$. Due to the logarithmic behavior, this breakdown happens at exponentially large distance for $f\lesssim M_\text{P}$. By contrast, it occurs instantaneously if $f\gtrsim M_\text{p}$.

\section{Singular string spacetimes}

The first exact solution of the Einstein equations for the exterior of an axionic string were given by Cohen and Kaplan (CK) \cite{ Cohen:1988sg}:

\be
\text{d}s^2=\frac{u}{u_0}(-\text{d}t^2+\text{d}z^2)+\gamma^2\left(\frac{u_0}{u}\right)^{1/2}\exp\left(\frac{u_0^2-u^2}{u_0}\right)(\text{d}u^2+\text{d}\theta^2).
\ee
Here $\gamma$ is an integration constant with the dimensions of length, $u_0$ is related to the axion decay constant\footnote{We use a slightly different normalization of the axion decay constant, hence the presence of an additional factor 2 compared to the original paper.} by $u_0 \equiv 2M_\text{P}^2/f^2$, and $0\le\theta<2\pi$. In these coordinates $u=\infty$ corresponds to the singular center of the string. At $u=0$, which corresponds to a cylindrical surface concentric with the string, one encounters another singularity. Both singularities are physical, as testified by the Kretschmann scalar 
\be
K=R^{\mu\nu\rho\sigma}R_{\mu\nu\rho\sigma}=\frac{\exp\left(\frac{2}{u_0}(u^2-u^2_0)\right)}{4\gamma^4u^3u_0^3}(32u^4-8u_0u^2+3u_0^2), \label{kretsch}
\ee
which diverges at these points. The proper distance between the outer singularity and the center is finite and reads
\be
r_\text{max}=\gamma\text{e}^{u_0/2}\left(\frac{u_0}{2}\right)^{5/8}\Gamma\left(\frac{3}{8}\right). \label{maxradius}
\ee
For small $f$ this distance is exponentially large and might not be physically relevant, e.g.~in a cosmological setting. However, for $f\gtrsim M_\text{P}$ one expects problems and we will discuss this momentarily.

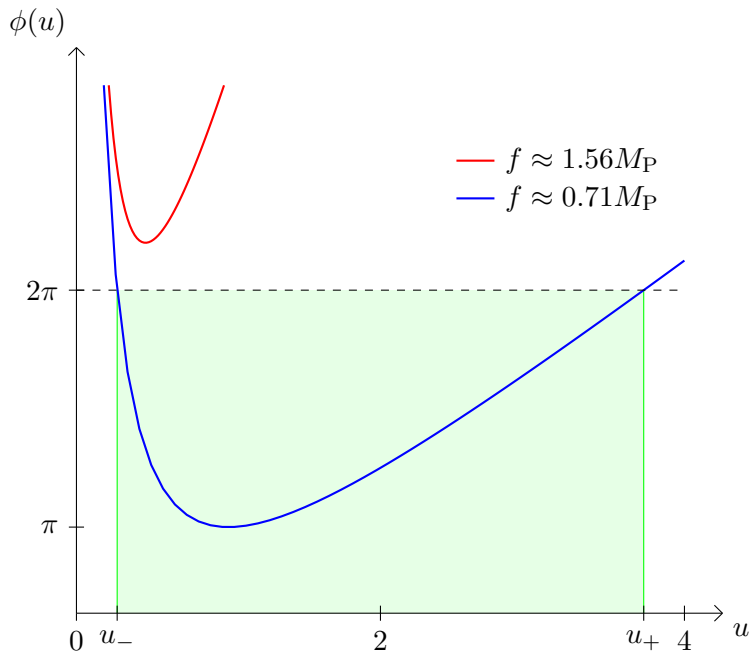
\begin{figure}[t]
\begin{center} 
\begin{tikzpicture}
\fill[green!10] (0.53589,2) -- (0.53589,6.28318) -- (7.4641,6.28318)--(7.4641,2) --cycle;
\draw[color=green] (0.53589,2) -- (0.53589,6.28318);
\draw[color=green] (7.4641,6.28318)--(7.4641,2);\
\draw (0.53589,1.9) node[anchor=north] {$u_-$} --(0.53589,2.1);
\draw (7.4641,1.9) node[anchor=north] {$u_+$} --(7.4641,2.1);
\draw (0,2) -- (8.5,2) node[anchor=north west] {$u$};
\draw (0,2) -- (0,9.5) node[anchor=south east] {$\phi(u)$};
\draw (8.5,2) -- +(-135:0.15);
\draw (8.5,2) -- +(135:0.15);
\draw (0,9.5) -- +(-45:0.15);
\draw (0,9.5) -- +(-135:0.15);
\draw (-0.1,6.28318) node[anchor=east] {$2\pi$} -- (0.1,6.28318);
\draw (-0.1,3.14159) node[anchor=east] {$\pi$} -- (0.1,3.14159);
\draw (0,1.9) node[anchor=north] {0} -- (0,2);
\draw (4,1.9) node[anchor=north] {2} -- (4,2.1);
\draw (8,1.9) node[anchor=north] {4} -- (8,2.1);
\draw[color=red,thick] (5,8) -- (5.5,8) node[anchor=west,color=black] {$f\approx 1.56M_\text{P}$};
\draw[color=blue,thick] (5,7.5) -- (5.5,7.5) node[anchor=west,color=black] {$f\approx 0.71M_\text{P}$};
\draw[color=blue,thick] plot[/tikz/domain=0.3604:8,/tikz/samples=50] (\x,{3.14159/2*(2/\x+\x/2)});
\draw[color=red,thick] plot[/tikz/domain=0.42556:1.94204,/tikz/samples=50] (\x,{3.14159*(1/\x+121/100*\x)});
\draw[dashed] (0,6.28318)--(8,6.28318);
\end{tikzpicture}
\caption{The angle $\phi$ defined in (\ref{angledef}) is shown as a function of the coordinate $u$ for \mbox{$f\approx 0.71M_\text{P}$} ($u_0\approx 4$) and $f\approx 1.56M_\text{P}$ ($u_0\approx0.83$). The shaded green area indicates the coordinate and angle range where one can visualize the space locally as conical, with a positive deficit angle (for $f\approx 0.71M_\text{P}$). $u=0$ corresponds to the outer singularity while the string center sits at $u=\infty$. The right intersection point of the blue and black dashed line at $u=u_+$ is roughly the core radius. For $f\approx 1.56M_\text{P}$ the angle is always larger than $2\pi$.}
\label{angle} 
\end{center}
\end{figure}

In order to gain some intuition for the CK spacetime we note that the region between $u+\text{d}u$ and $u$ at $t,z=\text{const.}$ describes an annulus. Its geometry approximates a piece of a cone and we find it convenient to denote the angle of the lateral surface\footnote{By this we mean the central angle of the corresponding circular segment.} of this cone by $\phi(u)$. In other words, $2\pi-\phi(u)$ is its deficit angle. A more formal definition is as follows: Consider a closed curve defined by $t,z,u=\text{const.}$ and parametrized by the coordinate $\theta$. We can determine the tangent vector $v$ to this curve at $\theta=0$ and parallel transport it along the curve until we reach $\theta=2\pi$, which is of course our starting point. Then calculate the angle between the parallel transported tangent vector $v'$ and $v$. This angle is $\phi(u)$. It turns out that $\phi=\text{d}C/\text{d}r$, where $C$ is the proper circumference and $r$ the proper radial distance. For the CK spacetime we find
\be
\phi(u)=\frac{\text{d}C}{\text{d}r}=2\pi\left(\frac{1}{4u}+\frac{u}{u_0}\right). \label{angledef}
\ee

For $f<\sqrt{2}M_\text{P}$ there is an interval of $u$ where $\phi(u)$ is increasing with $u$ and smaller than $2\pi$. Recall that increasing $u$ corresponds to decreasing distance to the string center (see Fig.~\ref{angle}). This is exactly the behavior we naively argued for at the end of Section 2. However, the angle is not bounded from above and there exist regions where it exceeds the critical value $2 \pi$. To be specific, we find $\phi > 2\pi$ for $u <u_{-}$ and $u > u_{+}$  where
\be
u_{\pm}=\frac{u_0}{2}\left(1\pm\sqrt{1-\frac{1}{u_0}}\right).
\ee
For $u_0\gg1$ one obtains $u_{+} \approx u_0$ and $u_{-} \approx 1/4$. We also see that for $u_0<1$ (which corresponds to $f > \sqrt{2}M_P$) the expressions $u_{\pm}$ become complex and the angle is strictly greater than $2\pi$ throughout the spacetime. We will come back to this point when assessing the existence of the CK solution. In any case, the minimum angle is reached at $u=\sqrt{u_0}/2$ and is given by $\phi_\text{min}=2\pi/\sqrt{u_0}$. This is illustrated in Fig.~\ref{angle}. 

We are now in a position to address the question under which circumstances the CK solution can be trusted and corresponds to a physically acceptable geometry. For example, we should discard solutions where curvature invariants are superplanckian everywhere. Thus a necessary condition for an acceptable solution is that there exist regions where the Kretschmann scalar $K$ is subplanckian.\footnote{Here we use the Kretschmann scalar instead of the scalar curvature $R$ since $K>R^2$ for all $u$. In particular, $R$ does not have a singularity at $u=0$ while $K$ does.} The minimal requirement for this to be possible is that $K$ is subplanckian at its minimum. This minimum lies at $u\approx\sqrt{u_0}/2$ and we arrive at the constraint
\be
\gamma\gtrsim\frac{1}{M_\text{P}}\frac{\text{e}^{-u_0/2}}{u_0^{5/8}}.\label{gammacons}
\ee
For $f\lesssim M_\text{P}$ the right-hand-side of this inequality is exponentially suppressed and hence a wide range of $\gamma$ gives rise to a weakly curved spacetime. However, also for $f\gtrsim M_\text{P}$ the choice of a sufficiently large $\gamma$ will ensure the existence of a weakly curved region of the CK spacetime. Eq. (\ref{gammacons}) also ensures that the radial size (\ref{maxradius}) of the spacetime obeys $r_\text{max}\gtrsim M_\text{P}^{-1}$. From the explicit form (\ref{kretsch}) of $K$ we see that we can make the region of small curvature between the inner and outer singularity always as large as desired by choosing $\gamma$ large enough. Pictorially, by increasing $\gamma$ we `stretch' the spacetime in the radial direction and thereby flatten its geometry.

Another criterion for deciding which solutions to trust employs the angle $\phi$. Although we are primarily interested in $f\gtrsim M_\text{P}$ let us examine for completeness first $f\le \sqrt{2}M_\text{P}$, i.e.~$u_0\ge1$. For an uncharged string we saw in Section 2 that the deficit angle is growing with the tension, i.e.~the corresponding angle $\phi$ decreases. By this we argued that the angle of the spacetime of a global string should decrease with the distance to the string center, which turned out to be correct at least far away from the outer singularity. This motivates us to conjecture that this behavior persists in every UV-completion of the core. Additionally, in order to avoid a conical singularity at the string center we expect every UV-completion to satisfy $\phi(\infty)=2\pi$. Combining this with our conjecture we can conclude that the angle of any UV-completion does not exceed $2\pi$. 

Such a UV-completion must somehow be matched onto the exterior CK spacetime at the core radius. Our upper bound on the angle in the core implies a lower bound for the core radius: $u_\text{core}\le u_+$. Recall that $u_+$ corresponds to the point where the angle would exceed $2\pi$ when approaching the string core, as indicated by the dashed line in Figure~\ref{angle}. For $f \leq \sqrt{2}M_P$ we find $u_+\sim u_0$ and hence $u_\text{core}\lesssim u_0$. Expressing this in the radial proper distance one finds $r_\text{core}\gtrsim\gamma$. The core radius can only take this minimal value if the Kretschmann scalar at $u\sim u_0$ is subplanckian, which yields a lower bound on $\gamma$: $\gamma\gtrsim f/M_\text{P}^2$.\footnote{This lower bound is stronger than (\ref{gammacons}) for $f\lesssim M_\text{P}$.} Altogether, there is no fundamental obstacle to the existence of a UV-completion of the string core for $f \leq \sqrt{2}M_P$.

The situation is completely different for $f>\sqrt{2}M_\text{P}$, i.e.~$u_0<1$. In this case we have $\phi>2\pi$ for all $u$. It is then not clear at all how the CK solution could be matched with a UV-completion of the core. A possible conclusion is that a UV-completion does not exist and hence the whole solution should be discarded. Recall that our main motivation is to gain a better understanding of the magnetic WGC for axions where the magnetically charged objects are given by strings. Our analysis in this section can thus be summarized as follows. For $f > \sqrt{2}M_\text{P}$ the CK string does not give rise to a trustworthy solution which could correspond to the magnetic object in the WGC for axions. This implies that either there are no string solutions for $f > \sqrt{2}M_\text{P}$, or string solutions exist but are not of CK type. We will explore the latter possibility in the next section. Note, however, that this conclusion crucially depends on the validity of our conjectured condition on the angle of UV-completions. This may well be too strong a requirement and if not true, CK solutions with $f>\sqrt{2}M_\text{P}$ might after all exist. For example, one may think of a higher dimensional UV-completion, where the requirement $\phi(\infty)=2\pi$ makes no sense.

Independently of this, even for $f\le\sqrt{2}M_\text{P}$, one  cannot accept this spacetime as physical due to the naked singularity in the exterior. It may however describe part of a cosmological spacetime or of a large string loop.

\section{Non-singular string spacetimes}

It turns out that it is possible to find a string spacetime which is not plagued by any physical singularity. Indeed, this can be done by allowing for a time-dependent metric and was first investigated by Gregory \cite{9606002}. This analysis uses a UV-completion of the axion model that is given by a complex scalar field $\Phi$ the phase of which plays the role of the axion:\footnote{In contrast to Gregory, we have canonically normalized the complex phase of $\Phi$.}
\be
\mathcal{L}=-\frac{1}{2}|\partial\Phi|^2-\frac{\lambda}{4}(|\Phi|^2-f^2)^2. \label{strpot}
\ee
The vacuum manifold is $\text{S}^1$ and gives rise to the string solutions we are interested in. Note that, although the metric depends on time, the field configuration is chosen to be static in Gregory's calculation, i.e.~the string has a constant width.

Let us describe the properties of this solution in more detail. Firstly, Gregory is able to show that the metric takes the form

\begin{equation}
\text{d}s^2=\text{e}^{A(r)}(-\text{d}t^2+\text{cosh}^2(\sqrt{b_0}t)\text{d}z^2)+\text{d}r^2+C^2(r)\text{d}\theta^2,
\end{equation}
where $b_0>0$ is a constant, and which in fact exhibits no singularities. We see that the spacetime of the string expands along the direction of the string while radial and angular metric components have no time-dependence. Furthermore, Gregory proves that this spacetime has a cosmological event horizon at finite proper distance from the string. This horizon encompasses the string and allows light to exit this interior space while it is not possible for physical objects to enter it from the exterior.

This event horizon moves inwards as $f$ increases. Therefore, the question arises what happens if the horizon enters the string core. Using analytic arguments, Gregory and Santos \cite{0208037} show that the string solution ceases to exist for $f>f_\text{max}$ with $6\le f_\text{max}^2/M_\text{P}^2\le 12$. We see two possible solutions that are left in the regime of parametrically large $f$: First, $\Phi=0$ together with a de Sitter metric is a classical solution. We discard it because of its instability. But second, and this will be the focus of the rest of this section, there is topological inflation.\footnote{Alexander Westphal informed us that, during one of his talks about \cite{1512.03768}, Andrei Linde also brought up the issue of using topological inflation in the context of the WGC for domain walls and axions.}

Topological inflation has been invented independently by Linde and Vilenkin \cite{9402031, 9402085} and occurs for topological defects whose width is comparable to the Hubble radius defined by the energy density in their core. Since the fields within the defects are at positive potential one expects the spacetime to be similar to de Sitter there. Hence, the defect expands exponentially in all directions. 

Let us apply the above mentioned condition for topological inflation to the string. In  order to do this assume that the UV-completion of the axion is given by the potential (\ref{strpot}). The axion is given by $\varphi\equiv \mbox{arg}(\Phi)$ and the cutoff of the model is set by the mass of the radial scalar $|\Phi|$:
\be
\Lambda^2\sim m^2_{|\Phi|} \sim \lambda f^2\,.
\ee
By balancing the gradient and potential energy density in the string core,
\be
\lambda f^4\stackrel{!}{\sim} (f/R)^2\,,
\ee 
the radius of the core is seen to be 
\be
R\sim \frac{1}{\sqrt{\lambda}\,f}\sim \frac{1}{m_{|\Phi|}}.
\ee
This also fixes the tension by integrating either gradient or potential energy density over the transverse section of the core:
\be
T\sim \lambda f^4 R^2 \sim (f/R)^2 R^2\sim f^2\,.
\ee
In fact, we see that the last result does not depend on the precise UV-completion since it can be derived more generally: Simply assume that, to realize a UV-completion, the $
\text{S}^1$ field space parametrized by $\varphi$ is embedded in a flat field space (in our case the $\mathbbm{R}^2$ parametrized by $\Phi$) with canonical euclidean metric. It is then clear that the field must cross a distance $\sim f$ in the string core of size $\sim R$. This implies a gradient energy density $\sim (f/R)^2$ and hence $T\sim f^2$, without any reference to the scalar potential and $\lambda$. 

Using this we can also state the condition for topological inflation to occur independently of the concrete potential. Let $V_0$ be the potential energy within the string which would be $V_0=(\lambda/4)f^4$ in the above example. Now, balancing the gradient energy density with the potential energy density leads to the estimate $R\sim f/\sqrt{V_0}$. The Hubble radius corresponding to the potential energy is given by $H_0^{-1}\sim M_\text{P}/\sqrt{V_0}$. Then, finally, the condition $R\gtrsim H_0^{-1}$ implies $f\gtrsim M_\text{P}$ which perfectly agrees with Gregory's result. Note, however, that our estimate has been derived without specifying the potential of the complex scalar explicitly. Only the embedding of the vacuum manifold $\text{S}^1$ into the field space $\mathbbm{R}^2$ of the UV-completion was assumed.

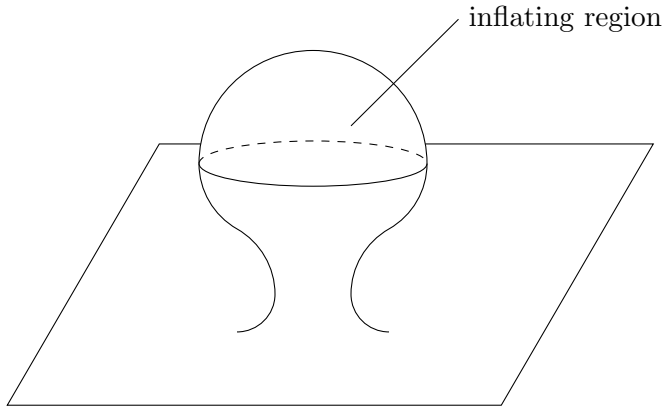
\begin{figure}
\begin{center} 
\begin{tikzpicture}
\draw (0,1.5) arc (90:0:1.5) arc (0:-60:1) arc (120:180:1) arc (180:270:0.5);
\draw (0,1.5) arc (90:180:1.5) arc (180:240:1) arc (60:0:1) arc (0:-90:0.5);
\draw (-1.5,0) arc [start angle=180, end angle=360, x radius=1.5, y radius=0.3];
\draw[dashed] (-1.5,0) arc [start angle=180, end angle=0, x radius=1.5, y radius=0.3];
\draw (10:1.5) -- ++(0:3) -- ++(-120:4) -- ++(180:6.5) -- ++(60:4) -- (170:1.5);
\draw (0.5,0.5) -- +(45:2) node[anchor=west] {inflating region};
\end{tikzpicture}
\caption{Embedding of a 2-dimensional slice of the spacetime for an inflating global monopole in Euclidean 3-space. The upper part of the `balloon' contains the inflating region.}
\label{balloon}
\end{center}
\end{figure}

In \cite{9804086} the model of Gregory was analyzed numerically for a general time-dependent metric and also allowing for time-dependent field configurations. It was found that a topological inflation scenario shows up for $f\gtrsim 0.23\cdot\sqrt{2}\cdot\sqrt{8\pi}M_\text{P}\approx 1.63M_\text{P}$. The spacetime structure of the analogue situation for a global monopole has been analyzed in \cite{9708005}.\footnote{Very recently the string spacetime structure has been studied numerically in \cite{1701.05572}. The authors have in particular studied models with $f<M_\text{P}$ and cosmic string solutions with multiple windings. They conclude that in such situations the outside observer can see a full transplanckian field cycle under certain conditions.} We expect this analysis to reflect at least the qualitative features of the string case. The main result is the presence of a horizon encompassing the monopole core that is analogous to the horizon in Gregory's string spacetime, i.e.~it is not possible to cross the horizon from the exterior while the opposite direction poses no problems. This horizon is the spacelike boundary of the inner inflating region. In fact, any observer that sits within this boundary will be expelled out of this region at some time as the monopole scalar field rolls down its potential and eventually oscillates about its minimum. These outer regions with the oscillating scalar field correspond to a matter dominated spacetime. During the evolution of the monopole the spacetime develops an inflating `balloon' which is connected by a throat to the exterior spacetime (cf.~Fig.~\ref{balloon}). For an observer it is possible to enter the throat as it contains not only the inflating region but also a matter dominated part.

Let us summarize what we have found so far. For $f\lesssim M_\text{P}$ there exist non-singular string spacetimes the metric of which is time-dependent while the field configuration is static. Such solutions are not available for $f\gtrsim M_\text{P}$. Instead, topological inflation becomes a possible scenario. However, the field providing the UV-completion of the axion then becomes time-dependent. 

Naively, one might want the magnetically charged object, i.e.~the string, to be static, at least in the sense that the field profile is static. According to this the spacetime found by Gregory is a well behaved UV-completion of the string for $f\lesssim M_\text{P}$. Whether topological inflation is a viable UV-completion of the string is an interesting open question. In any case, from the spacetime structure point of view it seems to be perfectly well behaved and is hence a UV-completion of a string with $f\gtrsim M_\text{P}$. Note however, that in both topological inflation and Gregory's solution a horizon is present which encompasses the string core and, contrary to the black hole horizon, shields the core from the exterior.

\section{Magnetic string in an effective theory with $f>M_\text{P}$}
In the following we present another candidate UV-completion for a magnetic string with $f>M_\text{P}$. To be specific, the theory of an axion with $f>M_\text{P}$ will be understood as an effective theory, which can be obtained from a more fundamental theory of two (or more) axions with subplanckian decay constants. Following \cite{1606.05552,1608.06951}, we then proceed to construct an effective string with $f>M_\text{P}$ out of the strings present in the more fundamental theory and examine it for potential pathologies. 

\subsection{Constructing the effective string with $f>M_\text{P}$}
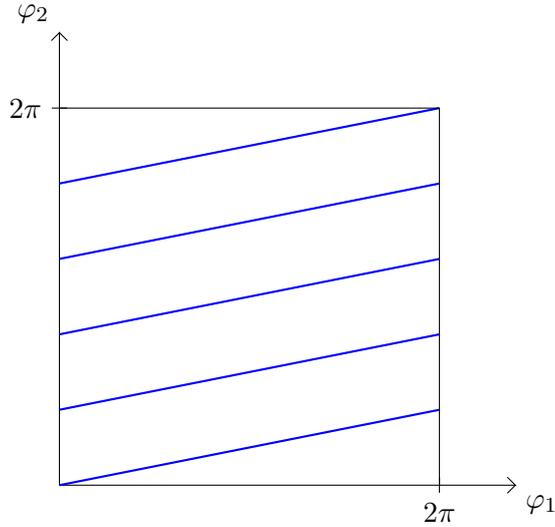
\begin{figure}
\begin{center} 
\begin{tikzpicture}
\draw (0,0) -- (6,0) node[anchor=north west] {$\varphi_1$} --+(135:0.15);
\draw (6,0) -- +(-135:0.15);
\draw (0,0) -- (0,6) node[anchor=south east] {$\varphi_2$} -- +(-135:0.15);
\draw (0,6) -- +(-45:0.15); 
\draw[thick,color=blue] (0,0) -- (5,1);
\draw[thick,color=blue] (0,1) -- (5,2);
\draw[thick,color=blue] (0,2) -- (5,3);
\draw[thick,color=blue] (0,3) -- (5,4);
\draw[thick,color=blue] (0,4) -- (5,5);
\draw (5,0) -- (5,5);
\draw (0,5) -- (5,5);
\draw (5,-0.1) node[anchor=north] {$2\pi$} -- (5,0.1);
\draw (-0.1,5) node[anchor=east]{$2\pi$} -- (0.1,5);
\end{tikzpicture}
\caption{Winding effective field space of total length $\sim Nf$ (shown for $N=5$).}
\label{winding} 
\end{center}
\end{figure}

We begin by recalling the UV completion of the last section:
\be
{\cal L}=-\frac{1}{2}|\partial\Phi|^2-\frac{\lambda}{4}(|\Phi|^2-f^2)^2\,.
\ee
Now consider the sum of two such Lagrangians, where we expand around the vacuum \mbox{$|\Phi_1|=f_1$,} $|\Phi_2|=f_2$. The axionic part can be written as:
\be
{\cal L}=-\frac{f_1^2}{2}(\partial\varphi_1)^2-\frac{f_2^2}{2}(\partial\varphi_2)^2
+\cdots\,.
\ee
For simplicity, we will take $f_1=f_2=f$ from now on. Famously, even if $f\ll M_P$, an effective axion with large decay constant can be obtained \cite{0409138}. The main idea behind \cite{0409138} is to design a potential on the $T^2$-field space parametrized by $\{\varphi_1,\varphi_2\}$ which forces the field onto a winding trajectory, e.g.
\be
\varphi_1=N\varphi_2\,,
\ee
with $N\gg 1$ (cf.~\cite{0912.1341} and Fig.~\ref{winding}). The desired potential can be viewed as arising from an appropriate combination of instantonic terms. 

As suggested in \cite{1503.07912}, a winding trajectory can alternatively be realized by an appropriate flux choice. This corresponds to making the orthogonal combination of axions massive by gauging \`a la Dvali \cite{0507215} (see also\cite{0811.1989,1101.0026}). This method can be viewed as a way of avoiding the 0-form WGC in the low-energy effective theory, as discussed more explicitly in \cite{1608.06951} in the 1-form context. As we will now see in detail, the flux of winding inflation \cite{1503.07912} is indeed the axion analogue of the effective field theory approach of \cite{1608.06951} for avoiding the magnetic WGC for vectors. However, we will also see that its success is more ambiguous than in the vector case.

Recall first the more familiar case of gauging of a shift-symmetric scalar by a 1-form\footnote{In fact, this has also been used inflationary model building in \cite{1503.01015}}
\be
{\cal L}=-\frac{f^2}{2}|d\varphi|^2 \qquad \to \qquad 
{\cal L}=-\frac{f^2}{2}|d\varphi+A_1|^2\,.
\ee
Equivalently, with $F_1=d\varphi$, this reads
\be
{\cal L}=-\frac{f^2}{2}|F_1|^2 \qquad \to \qquad 
{\cal L}=-\frac{f^2}{2}|F_1+A_1|^2\,.\label{0fg}
\ee
By analogy, we can write down the lagrangian of a 3-form/$(-1)$-form theory as 
\be
{\cal L_{KS}}=-\frac{1}{2g^2}|F_0|^2\,,
\ee
where $F_0$ is a dimensionless field strength, quantized in units of $2\pi$.
The coupling constant $g$ has mass dimension $(-2)$. Then, in analogy to (\ref{0fg}), we gauge this theory according to
\be
{\cal L_{KS}}=-\frac{1}{2g^2}|F_0|^2\qquad\to\qquad 
{\cal L_{KS}}=-\frac{1}{2g^2}|F_0+\varphi|^2\,.
\ee
In other words, we simply add our 0-form potential to the field strength in the kinetic term of the ungauged model. This is just a simplified version of the idea in \cite{0507215}, which in the context of inflation is also known as the `Kaloper-Sorbo' approach to axion monodromies \cite{0811.1989, 1101.0026}. Indeed, we simply skipped the detour via the 2-form theory dual to the axion.

In our context, it is interesting to gauge the combination $\varphi_1-N\varphi_2$, i.e.~we consider the model defined by
\be
{\cal L}=-\frac{f^2}{2}(\partial\varphi_1)^2-\frac{f^2}{2}(\partial\varphi_2)^2
-\frac{1}{2g^2}|F_0+\varphi_1-N\varphi_2|^2+\cdots\, . \label{pot} 
\ee
In addition, we include the various charged objects (two types of strings and one type of domain wall) and, if desired, the UV-completion of the axions discussed above.
For definiteness, we focus on the background with $F_0=0$. Clearly, the field space of the effective axion is the submanifold of $\text{T}^2$ specified by the equation $\varphi_2=\varphi_1/N$ (cf.~Fig.~\ref{winding}).

Let us make a brief detour to explain what happens at the conceptual level:
Originally, we have a field space parameterized as $\{\varphi_1,\varphi_2\}$, with the discrete symmetry group ${\mathbbm Z}^2$ (shifts by $2\pi\,\{m,n\}$) being gauged. The $F_0$ theory is unrelated. Then, we associate with any of the ${\mathbbm Z}^2$ shifts a particular transformation of $F_0$, in our case 
\be
F_0\quad\to\quad F_0\,-\,2\pi (m-Nn)\,.\label{mnf0}
\ee
One could say that we picked a group homomorphism from ${\mathbbm Z}^2$ to the proposed ${\mathbbm Z}$ symmetry of $F_0$. Now, $|F_0|^2$ is not an invariant Lagrangian any more, but (\ref{pot}) provides the appropriately modified, invariant  version. This Lagrangian necessarily couples the $F_0$ theory with the $\{\varphi_1,\varphi_2\}$ theory.

To continue, let us introduce the alternative field basis
\bea
\psi&=&N\varphi_1+\,\,\varphi_2\\
\chi&=&\,\,\varphi_1-N\varphi_2\, ,
\eea
such that
\be
{\cal L}=-\frac{f^2}{2(N^2+1)}\big[(\partial\psi)^2+(\partial\chi)^2\big]
-\frac{1}{2g^2}|F_0+\chi|^2+\cdots\,. \label{modlag}
\ee
We identify $\psi$ as the effective low-energy axion which is a periodic field with decay constant $f\sqrt{N^2+1}$. By contrast, $\chi$ is massive and, in addition, closed loops in the field space of $\chi$ are only possible at the expense of passing a domain wall of the $F_0$ theory.

\begin{figure}[t]
\begin{center} 
\begin{tikzpicture}
\draw[color=green,thick] (0,0) -- (3,0);
\draw[color=green,thick,dotted] (3,0) -- (4,0);
\fill[color=blue] (0,0) circle (0.1);
\draw[thick] (0,0) circle (1);
\draw[thick] (0,0) ++ (0,-1) -- +(135:0.2);
\draw[thick] (0,0) ++ (0,-1) -- +(-135:0.2);
\draw[thick] (-1.2,0) node[anchor=south east] {$2\pi$} node[anchor= north east] {$0$} -- (-0.8,0);
\draw (0,0) +(135:2) node {$\varphi_1$};

\draw[thick] (0,0) +(-10:2.5) node[anchor=north] {$F_0$ }arc (-10:10:2.5) node[anchor=south] {$F_0-2\pi$};
\draw[thick] (0,0) ++(10:2.5) -- +(-130:0.15);
\draw[thick] (0,0) ++(10:2.5) -- +(-40:0.15); 
\end{tikzpicture}
\caption{The string type 1 is shown with one domain wall attached to it. The jump of $F_0$ across the wall is indicated.}
\label{string1} 
\end{center}
\end{figure}
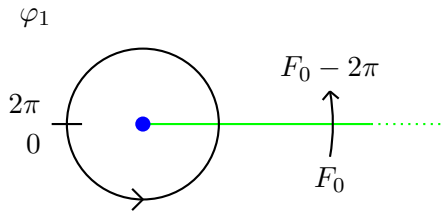

Let us make this last point about domain walls more explicit. For this purpose recall that a string is present whenever the axion follows a closed loop in its field space. Since we have two axions here, $\varphi_1$ and $\varphi_2$, we also have two different species of strings, called string type 1 and 2 from now on. For example, integrating the field strength $\text{d}\varphi_1$ along a closed path in space which encompasses a string we must have 
\be
\oint\text{d}\varphi_1=2\pi.
\ee
If this loop does not contain an additional string of type 2 the corresponding integral of $\varphi_2$ vanishes. Therefore, we have the general relation
\be
\oint\text{d}\varphi_i= 2\pi N_i\quad i=1,2\quad ,
\ee
where $N_i$ denotes the effective number of strings of type $i$ that lie within the closed integration path. Similarly, a domain wall is defined by the set of points in spacetime at which the value of $F_0$ jumps by an amount that is determined by the unit of quantization of $F_0$, in our case $2\pi$. Combining these facts with the gauging procedure described above we find a relation between strings and domain walls, which we will describe below.

\begin{figure}[t]
\begin{center} 
\begin{tikzpicture}
\draw[color=green,thick] (0,0) -- (3,0);
\draw[color=green,thick,dotted] (3,0) -- (4,0);

\draw[color=green,thick] (0,0) -- +(45:3);
\draw[color=green,thick,dotted] (0,0) ++(45:3) -- (45:4);

\draw[color=green,thick] (0,0) -- +(90:3);
\draw[color=green,thick,dotted] (0,0) ++(90:3) -- (90:4);

\draw[color=green,thick] (0,0) -- +(-45:3);
\draw[color=green,thick,dotted] (0,0) ++(-45:3) -- (-45:4);

\fill[color=red] (0,0) circle (0.1);
\draw[thick] (0,0) circle (1);
\draw[thick] (0,0) ++ (0,-1) -- +(135:0.2);
\draw[thick] (0,0) ++ (0,-1) -- +(-135:0.2);
\draw[thick] (-1.2,0) node[anchor=south east] {$2\pi$} node[anchor= north east] {$0$} -- (-0.8,0);
\draw (0,0) +(135:2) node {$\varphi_2$};

\draw[thick] (0,0) +(-60:2.5) node[anchor=east] {$F_0-2\pi$}arc (-60:-30:2.5);
\draw[thick] (0,0) ++(-30:2.5) -- +(-165:0.15);
\draw[thick] (0,0) ++(-30:2.5) -- +(-75:0.15); 

\draw[thick] (0,0) +(-15:2.5) node[anchor=north] {$F_0$ }arc (-15:15:2.5);
\draw[thick] (0,0) ++(15:2.5) -- +(-120:0.15);
\draw[thick] (0,0) ++(15:2.5) -- +(-30:0.15); 

\draw[thick] (0,0) +(30:2.5) node[anchor=north] {$F_0+2\pi$ }arc (30:60:2.5);
\draw[thick] (0,0) ++(60:2.5) -- +(-75:0.15);
\draw[thick] (0,0) ++(60:2.5) -- +(15:0.15); 

\draw[thick] (0,0) +(75:2.5) node[anchor=west] {$F_0+4\pi$ }arc (75:105:2.5);
\draw[thick] (0,0) ++(105:2.5) -- +(-30:0.15);
\draw[thick] (0,0) ++(105:2.5) -- +(60:0.15);
\draw[thick,dotted] (0,0) ++(110:2.5) arc (110:121:2.5);

\end{tikzpicture}
\caption{The string type 2 is shown with $N$ domain walls attached to it. The jumps of $F_0$ across the walls are indicated.}
\label{string2} 
\end{center}
\end{figure}
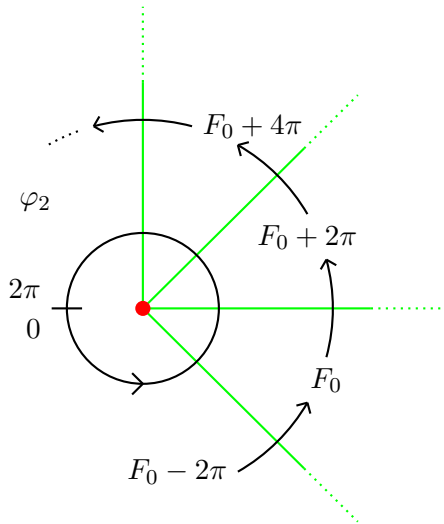

Consider a single string of type 1 embedded in a background of constant $F_0$. Going once around this string in space corresponds to starting with a value $\varphi_1=0$ at some point $P$ and finally reaching $\varphi_1=2\pi$ when closing the path in space at $P$ again. Since there is no string of type 2 present, $\varphi_2$ takes on the same value when we return to $P$. Continuity demands this to be the same field configuration up to gauge equivalence. In order to identify the final configuration ($\varphi_1=2\pi$) with the initial one ($\varphi_1=0$) we can perform a gauge transformation $\varphi_1\rightarrow\varphi_1-2\pi$. However, according to (\ref{mnf0}), this necessarily implies a shift $F_0\rightarrow F_0+2\pi$ which is not the same field configuration we had in the beginning. To make this consistent there has to exist a domain wall attached to the type 1 string such that $F_0\rightarrow F_0-2\pi$ when one crosses this wall while circumnavigating the string (cf.~Fig.~\ref{string1}). Then the above gauge transformation gives us back the initial field configuration. We can repeat the argument for a string of type 2. The result is that a type 2 string must be attached to $N$ domain walls as indicated in Fig.~\ref{string2}. This follows straightforwardly from the gauge transformation of $F_0$ \eqref{mnf0}.\footnote{A similar situation with strings ending on monopoles occurs in models of semilocal strings \cite{Vachaspati:1991dz}.}

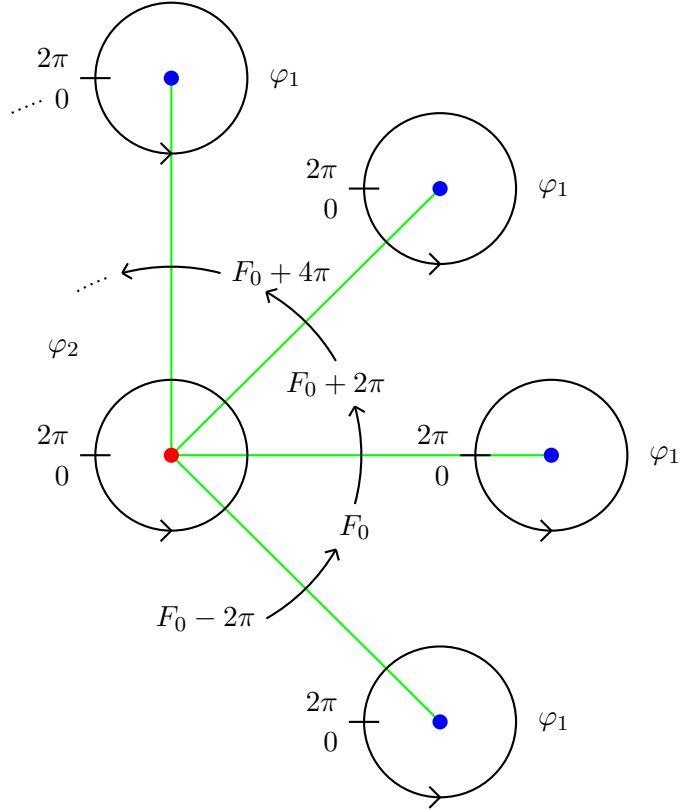
\begin{figure}[t]
\begin{center} 
\begin{tikzpicture}
\draw[color=green,thick] (0,0) -- (5,0);

\draw[color=green,thick] (0,0) -- +(45:5);

\draw[color=green,thick] (0,0) -- +(90:5);

\draw[color=green,thick] (0,0) -- +(-45:5);

\fill[color=red] (0,0) circle (0.1);
\draw[thick] (0,0) circle (1);
\draw[thick] (0,0) ++ (0,-1) -- +(135:0.2);
\draw[thick] (0,0) ++ (0,-1) -- +(-135:0.2);
\draw[thick] (-1.2,0) node[anchor=south east] {$2\pi$} node[anchor= north east] {$0$} -- (-0.8,0);
\draw (0,0) +(135:2) node {$\varphi_2$};

\draw[thick] (0,0) +(-60:2.5) node[anchor=east] {$F_0-2\pi$}arc (-60:-30:2.5);
\draw[thick] (0,0) ++(-30:2.5) -- +(-165:0.15);
\draw[thick] (0,0) ++(-30:2.5) -- +(-75:0.15); 

\draw[thick] (0,0) +(-15:2.5) node[anchor=north] {$F_0$ }arc (-15:15:2.5);
\draw[thick] (0,0) ++(15:2.5) -- +(-120:0.15);
\draw[thick] (0,0) ++(15:2.5) -- +(-30:0.15); 

\draw[thick] (0,0) +(30:2.5) node[anchor=north] {$F_0+2\pi$ }arc (30:60:2.5);
\draw[thick] (0,0) ++(60:2.5) -- +(-75:0.15);
\draw[thick] (0,0) ++(60:2.5) -- +(15:0.15); 

\draw[thick] (0,0) +(75:2.5) node[anchor=west] {$F_0+4\pi$ }arc (75:105:2.5);
\draw[thick] (0,0) ++(105:2.5) -- +(-30:0.15);
\draw[thick] (0,0) ++(105:2.5) -- +(60:0.15);
\draw[thick,dotted] (0,0) ++(110:2.5) arc (110:121:2.5);
\draw[thick,dotted] (0,0) ++(110:5) arc (110:115:5);

\fill[color=blue] (0,0) ++(-45:5) circle (0.1);
\draw[thick] (0,0) ++(-45:5) circle (1);
\draw[thick] (0,0) ++(-45:5) ++ (0,-1) -- +(135:0.2);
\draw[thick] (0,0) ++(-45:5) ++ (0,-1) -- +(-135:0.2);
\draw[thick] (0,0) ++(-45:5) ++ (-1.2,0) node[anchor=south east] {$2\pi$} node[anchor= north east] {$0$} -- +(0.4,0);
\draw[thick] (0,0) ++(-45:5) ++(1.5,0) node {$\varphi_1$};

\fill[color=blue] (0,0) ++(0:5) circle (0.1);
\draw[thick] (0,0) ++(0:5) circle (1);
\draw[thick] (0,0) ++(0:5) ++ (0,-1) -- +(135:0.2);
\draw[thick] (0,0) ++(0:5) ++ (0,-1) -- +(-135:0.2);
\draw[thick] (0,0) ++(0:5) ++ (-1.2,0) node[anchor=south east] {$2\pi$} node[anchor= north east] {$0$} -- +(0.4,0);
\draw[thick] (0,0) ++(0:5) ++(1.5,0) node {$\varphi_1$};

\fill[color=blue] (0,0) ++(45:5) circle (0.1);
\draw[thick] (0,0) ++(45:5) circle (1);
\draw[thick] (0,0) ++(45:5) ++ (0,-1) -- +(135:0.2);
\draw[thick] (0,0) ++(45:5) ++ (0,-1) -- +(-135:0.2);
\draw[thick] (0,0) ++(45:5) ++ (-1.2,0) node[anchor=south east] {$2\pi$} node[anchor= north east] {$0$} -- +(0.4,0);
\draw[thick] (0,0) ++(45:5) ++(1.5,0) node {$\varphi_1$};

\fill[color=blue] (0,0) ++(90:5) circle (0.1);
\draw[thick] (0,0) ++(90:5) circle (1);
\draw[thick] (0,0) ++(90:5) ++ (0,-1) -- +(135:0.2);
\draw[thick] (0,0) ++(90:5) ++ (0,-1) -- +(-135:0.2);
\draw[thick] (0,0) ++(90:5) ++ (-1.2,0) node[anchor=south east] {$2\pi$} node[anchor= north east] {$0$} -- +(0.4,0);
\draw[thick] (0,0) ++(90:5) ++(1.5,0) node {$\varphi_1$};

\end{tikzpicture}
\caption{The unique combination of the building blocks given by string type 1 and 2 gives rise to an urchin-like structure shown in this figure consisting of one central string of type 2 connected to $N$ strings of type 1 by $N$ domain walls. The orientation of the axions and the jumps of $F_0$ are shown.}
\label{urchin} 
\end{center}
\end{figure}

We are now in a position to build an effective string out of strings of type 1 and 2 connected by domain walls. This construction is parallel to the procedures in \cite{1606.05552,1608.06951}. In particular, take one string of type 2 and attach one string type 1 to each free end of the $N$ domain walls that come with the string type 2 (cf.~Fig.~\ref{urchin}). We observe that the orientation of the domain walls automatically fits the shifts in $F_0$ as discussed for the two types of strings separately. The resulting object is a string that is uncharged under the massive field $\chi$ in the sense that $\chi\rightarrow\chi$ when going once around it. Similarly we have $\psi\rightarrow\psi+2\pi(N^2+1)$. Hence, the effective string just constructed is the string corresponding to the axion $\psi$ of the low energy theory with decay constant $f_\text{eff}=f\sqrt{N^2+1}$. The massive axion $\chi$ is not visible outside this effective string, i.e.~at low energies.

The important point for us is that this object corresponds to a microscopic construction of a string for $f > M_P$. In the following we wish to examine whether this effective string can exist as a field-theoretical object. For example, we will only trust this object without gravitational backreaction if the effective string tension is subplanckian, i.e.~$T_{\textrm{eff}} \lesssim M_P^2$. At the same time, we have seen in the previous sections that string solutions with $f > M_P$ can at most exist as dynamical solutions giving rise to topological inflation. Hence, we also wish to determine to what extent our effective string is consistent with these previous findings.

\subsection{Estimating the string tension}
Let us therefore estimate the tension of the effective string. All in all, there are three contributions to this: 
\begin{itemize}
\item The tension of the individual elementary strings contained in the effective string.
\item The tension generated by the mutual interaction of the elementary strings.
\item The tension due to the domain walls.
\end{itemize}

Elementary strings have tensions $f^2$ which sum up to a total of $(N+1)f^2\sim f_\text{eff}^2/N$. This can be kept subplanckian by choosing $N$ large enough (for fixed $f_{\textrm{eff}}$) and hence poses no problem for us.

Next consider the mutual interaction of the elementary strings. This contribution is very complicated to determine as it depends on the detailed configuration of the $N+1$ elementary strings connected by $N$ domain walls. However, one can make the following approximation: Let $R$ denote the radius of the effective string. The average density of strings in the cross section of the effective string is given by $\sim N/R^2$. This corresponds to an average distance $\sim R/\sqrt{N}$ between single strings. For large $N$ this is certainly much smaller than the size $R$ of the effective string and it is reasonable to describe the elementary string distribution in terms of a continuous charge density $\varrho$, i.e.~the charge per cross section area. In addition, one certainly expects the $N$ strings of type 1 to arrange themselves approximately symmetrically around the single string of type 2. Hence, in the continuum limit, we argue that the charge density depends only on the distance $r$ from the center of the effective string and we can write $\varrho=\varrho(r)$. This density is of course normalized by the charge of the effective string, i.e. 
\be
\int_0^{2\pi}\text{d}\varphi\int_0^Rr\text{d}r\varrho(r)=N+1.
\ee
Define
\be
Q(r)=\int_0^{2\pi}\text{d}\varphi\int_0^rr'\text{d}r'\varrho(r')
\ee
to be the string charge contained in a cylinder of radius $r$ centered on the inner type 2 string. The norm of the field strength of such a charge distribution at radius $r$ is given by $Q(r)/r$, while the surrounding charge does not generate a contribution to the field, which is completely analogous to classical Electrodynamics. Exterior to our effective string, i.e.~for $r>R$, we have of course \mbox{$Q(r)=N+1$.} The correct $r$-dependence of $\varrho$ might be calculated by balancing the forces on a elementary string at distance $r$ from the center due to the attached domain wall and the field of all string charge contained in the cylinder of radius $r$. The repulsive force per length between two strings having positive charges $n_1$ and $n_2$ respectively is given by $2\pi n_1n_2f^2/r$ with $r$ being the distance between them. The attractive fore per length between a string type 1 and the central string type 2 due to the connecting domain wall is given by the tension $W$ of the domain wall. With this information we can write down the balance equation for the forces,
\be
W\sim\frac{Q(r)f^2}{r},
\ee
which results in $Q(r)\sim(W)r/f^2$. The normalization $Q(R)=N+1$ determines the string radius $R$ as 
\be
R\sim Nf^2/W. \label{stringradius}
\ee

Now we can calculate the effective tension due to the mutual interaction of the elementary strings. It is given by integrating the energy density over the cross section of the effective string, i.e.

\be
\int_0^{2\pi}\text{d}\varphi\int_0^Rr\text{d}r\frac{f^2}{2}\frac{Q(r)^2}{r^2}\sim N^2f^2\sim f_\text{eff}^2.
\ee
This shows that the effective tension generically becomes superplanckian if $f_\text{eff}$ is chosen superplanckian.

Finally, we determine the contribution of the tension of the domain walls. The tension due to one domain wall is $\sim RW$. Collecting the contribution of each wall in the effective string gives a tension $\sim NRW$.

In addition to this inherent domain wall tension, there is another contribution due to the excitation of the massive field when crossing a domain wall. From (\ref{modlag}) we see that $\chi=-F_0$ in the vacuum. Since a domain wall is accompanied by a jump in $F_0$, it is not possible for $\chi$ to stay at the potential minimum and it is hence excited when crossing a domain wall. This is an additional contribution to the tension of the domain wall and can be estimated as follows. Consider a domain wall at a point far away from any strings that may be attached to it. Then, the translational symmetry tangent to the wall implies that the massive field depends only on the direction orthogonal to the wall, the coordinate of which may be chosen to be $x$. Let the background field be $F_0=F_-$ for $x<0$ and $F_0=F_+$ for $x>0$ and define $\Delta F_0=F_+-F_-=2\pi$, as appropriate for a single domain wall. Far away from the wall we want the massive field $\chi$ to occupy the vacuum, i.e.~we demand $\lim_{x\rightarrow -\infty}\chi(x)=-F_-$ and $\lim_{x\rightarrow +\infty}\chi(x)=-F_+$. The equations of motion for the Lagrangian (\ref{modlag}) with these boundary conditions are solved by

\be
\chi(x)=
\begin{cases}
-\pi\text{e}^{\alpha x}-F_-\quad\text{for}\quad x<0\\
\pi\text{e}^{-\alpha x}-F_+\quad\;\text{for}\quad x\ge 0\\
\end{cases}
,
\ee
where $\alpha=\sqrt{N^2+1}/(gf)$. The typical width of this field profile is given by $\alpha^{-1}$. The field profile is expected to be modified if one approaches the string at the end of the wall to a distance of less than $\alpha^{-1}$. This should be irrelevant to our calculation if the typical wall length $R$ obeys $R\gg \alpha^{-1}$. Since the deviation from the background $F_0$ is always $\mathcal{O}(1)$ the potential energy density of the above field configuration is \mbox{$\sim g^{-2}$.} Integrating this over the field profile with typical width $\alpha^{-1}$ yields a contribution \mbox{$W_\chi\sim f/(gN)$} to the tension of the domain wall. Thus, the overall contribution to the tension of the effective string reads $\sim(NW+f/g)R \sim (NW+f_\text{eff}/(Ng))R$. Using the expression (\ref{stringradius})\footnote{Note that now we have to use the full domain wall tension given by $W+W_\chi$ in this formula.} for the radius $R$ we find $\sim f_\text{eff}^2$ for the total domain wall contribution to the string tension, which is the same as for the contribution from the mutual string interaction.

We are therefore forced to conclude that the existence of strings with superplanckian charge and subplanckian tension is very questionable. Even if we had found that it is possible, we were faced with the problem of the badly behaving spacetime of such a string when including gravity. This was extensively discussed in the previous two sections. There we presented the work of Gregory who argued that abandoning the requirement of a static metric allows for a topological inflation scenario for strings with superplanckian charge. Hence, Saraswat's method for constructing effective charged objects allowed us to find a possible UV-completion of topological inflation in terms of a composite string with superplanckian charge.

\section{A comment on the field range of pseudo-axions}

While this paper mostly discusses the field range of axions, our interest is of course more generally in all scalar fields with potentially large field displacement, including non-periodic ones. We want to discuss a potential loophole arising in this context \cite{0804.3961} and how it may be closed due to a stringy back-reaction effect.

Indeed, consider the schematic 10d type-IIA or IIB string-frame action (with $l_s=1$)
\be
S\sim\frac{1}{g_\text{s}^2}\int \text{d}^{10}x\,\sqrt{g}\,{\cal R}+\frac{1}{g_\text{s}}\int_{\text{D}p}\text{d}^{p+1}x\,\sqrt{g}\,,
\ee
with ${\cal R}$ the Ricci scalar, $g_\text{s}=\exp(\phi)$ the string coupling, and 
$g$ the determinant of the metric or induced metric. We ignore all factors of $2\pi$ etc.~since they are not essential to make our point.

Considering specifically type-IIA theory with D8-branes and compactifying to 5d on a $\text{T}^5$ with radius $\sim 1$ we find
\be
S\sim \frac{1}{g_\text{s}^2}\int \text{d}^5x\,\sqrt{g}\,{\cal R}+\frac{1}{g_\text{s}}\int \text{d}^4x\,\sqrt{g}\,.
\ee
Here we assumed that the D8 wraps the compact torus. Now, compactifying further on an $\text{S}^1$ of radius $R$ gives
\be
S\sim \frac{2\pi R}{g_\text{s}^2}\int \text{d}^4x\,\sqrt{g}\,{\cal R}+\int\text{d}^4x\frac{R^2}{g_\text{s}}(\partial\varphi)^2+...\,,
\ee
where $\varphi\in[0,2\pi)$ is the scalar characterizing the brane-position within the $\text{S}^1$. First, we immediately see that
\be
M_\text{P}\sim \sqrt{R}/g_\text{s}\qquad \mbox{and}\qquad f\sim R/\sqrt{g_\text{s}}\,,
\ee
such that 
\be
f/M_\text{P}\sim \sqrt{g_\text{s}R}\label{fmp}
\ee
can apparently easily be made parametrically large. However, $\varphi$ is not a proper axion. Indeed, to ensure tadpole cancellation one has to replace $\text{S}^1$ by $\text{S}^1/\text{Z}_2$, with $\text{O}8$-planes at both ends of the interval and a total of $16$ D8 branes (i.e.~there are actually 16 scalars of the type of $\varphi$). The field space of $\varphi$ is now the orbifold $\text{S}^1/\text{Z}_2$ (hence `pseudo-axion'), but its size is still $\sim f$ and potentially large in 4d Planck units.

Such effectively one-dimensional compact spaces are of course a well-known exception to the no-go theorem of \cite{0105204} (which says that the field-space of transverse brane-motion is, in general, subplanckian). Indeed, a model of potentially transplanckian brane inflation making use of this possibility was suggested in \cite{0804.3961}\footnote{
It 
actually uses type-IIB on K3$\times(\text{T}^2/\text{Z}_2)$ with D7 and D3 branes, considering a limit where the pillow-shaped space $\text{T}^2/\text{Z}_2$ is extremely prolonged. This is clearly related by $T$-duality to our toy-model if one were to replace the K3 by a $\text{T}^4$.
}.
Establishing such a model is associated with all the usual complications, like creating a non-zero potential and moduli stabilization. Here, we are not interested in this. For us, just having the underlying SUSY-Minkowski model with a large field range would be interesting.

Now, unfortunately, in our D8-brane toy model this possibility is censored in a very simple manner which, one might suspect, will carry over to any related effectively one-dimensional compact geometry. Indeed, the branes are effective domain walls and their field strength is a top-form-flux. The field strength does not decay with growing distance from the brane, leading to strong back-reaction. As worked out in detail e.g.~in \cite{9601150,0103233}, the metric becomes singular if one of the 8 D8-branes which have to sit on each O8-plane is moved away too far. Specifically, metric coefficients and dilaton profile behave as powers of
\be
H=1-g_\text{s}x\cdot\,\rm const.\,,
\ee
such that the field range of $\varphi$ is limited by 
\be
\varphi_{\rm max}\sim x_{\rm max}/R \sim 1/(g_sR)\,.
\ee
Here $x$ is the coordinate of the orbifold $\text{S}^1/\text{Z}_2$. This bound is of course only relevant if $1/(g_\text{s}R)<1$. With this, the canonical field range limit due to back reaction, in Planck units, becomes
\be
\frac{f\varphi_{\rm max}}{M_\text{P}}\sim \frac{1}{\sqrt{g_\text{s} R}}\,.
\label{brf}
\ee

To summarize, our interval limited by two O8-planes can in principle be made large as long as the branes cancel the tadpole locally. The naive field range due to brane motion on this interval is $\sim \sqrt{g_\text{s}R}$ and can also become large. Displacing the branes from the O-plane corresponds to a perfectly flat 4d directions due to unbroken ${\cal N}=4$ supersymmetry. However, once one actually starts moving one of the branes off an O-plane and along the interval, the geometry breaks down long before one reaches the O-plane at the opposite end. Indeed, according to (\ref{brf}) this breakdown happens at the  field displacement $1/\sqrt{g_\text{s}R}$. But this number is small since, according to (\ref{fmp}), we need $\sqrt{g_\text{s}}R$ to be large to have any hope for transplanckian field ranges to begin with. The optimum is achieved by choosing $(g_\text{s}R)\sim 1$, which allows for Planck-scale field displacement, but not more than that.

Because of the high amount of supersymmetry there is no potential barrier and we expect that the breakdown at $\varphi\sim\varphi_{\rm max}$ must be due to an exponential decrease of the cutoff-scale. This is expected in the context of the swampland conjecture \cite{0605264} (see also \cite{0802.3923}) and has recently been further discussed in \cite{1610.00010}. It may be worthwhile understanding this connection in more detail. We find the apparent possibility of a transplanckian field range and the way in which it finally fails interesting for the following reason: One might have thought that, for a pseudo-axion the whole logic of the WGC is irrelevant and the field range can become large. This appears to be borne out by branes on $\text{S}^1/\text{Z}_2$ as explained, but fails due to a relatively subtle stringy reason. Thus, there may be something deeper and more mysterious than either the WGC or the decompactification logic emphasized in \cite{0605264} which is {\it really} behind the problems with large field ranges and large-field inflation.

\section{Conclusions}

The two standard arguments against superplanckian axion decay constants, $f\gtrsim M_\text{P}$, are the loss of control over instantonic corrections (the electric WGC) and problems with stringy realizations. In this paper, we have tried to argue for the same conclusion using the magnetic WGC, i.e. the requirement that a dual charged object (a string) exists in the effective theory.

First, we wrote down the general formula for the cutoff resulting from the magnetic WGC and attempted an extrapolation to the naively singular case of axions. This suggests that $f\gtrsim M_\text{P}$ should be forbidden.

Second, we examined explicit string solutions with $f\gtrsim M_\text{P}$, looking for possible inconsistencies. Static solutions of this type (the Cohen-Kaplan spacetime of an axionic charged string) are singular in both the UV and the IR. Accepting the IR problems as a special feature of infinitely extended strings in general relativity, the crucial question is whether the UV problems can be cured by an appropriate UV-completion. We argued that smooth, 4d UV-completions can not lead to negative deficit angles and showed that this excludes static string spacetimes with $f\gtrsim M_\text{P}$. Superplanckian axion decay constants would then be ruled out.

We then relaxed the requirement of time-independence in two steps. First, one can allow for a time-dependent metric while the field configuration should still be static, i.e.~the string width is fixed. Such solutions, found by Gregory, can be singularity-free in the IR and have a smooth UV-completion. They possess a horizon encompassing the string core. However, they exist only  for $f\lesssim M_\text{P}$.

Next, we considered admitting general time-dependent field configurations. In this case, the set-up of topological inflation provides a UV-completion of the string, even for $f\gtrsim M_\text{P}$. In fact, topological inflation requires superplanckian $f$. We argued that the corresponding spacetime has a horizon, similar to that of Gegory's spacetime for $f\lesssim M_\text{P}$. There are two reasons why one might want to reject topological inflation as a string solution in the sense of the magnetic WGC: One is the time-dependence which one could consider unnatural for an object that is supposed to be the analogue of a magnetic monopole. The second is the presence of a horizon (although the latter is very different from the black hole horizon potentially hiding a magnetic monopole). However, consistency would then force us to reject Gregory's solution for $f\lesssim M_\text{P}$ as well, such that no string solution is acceptable at all. In other words: If we demand that string solutions with $f\lesssim M_\text{P}$
exist, we cannot use the horizon as an argument against topological inflation. The interpretation of these observations remains open.

Finally, we tried to apply Saraswat's recent observation \cite{1608.06951} that certain low-energy effective theories can avoid the collapse of the minimally charged monopole to a black hole, even though they violate the WGC. We considered theories with two subplanckian axions and two types of strings. At low energies such models can have one effective superplanckian axion coupled to one effective string. However, in contrast to the gauge-theory case, the tension of this string remains superplanckian. Although we do not provide an exact solution of this system including gravity, the previously discussed properties of string spacetimes suggest very strongly that time dependence will be an unavoidable feature. But it may be, that this time-dependence is restricted to the metric while corresponding field configurations remain static, as is the case in Gregory's solution. It would be important to understand this in detail.

The decisive question whether the magnetic WGC rules out superplanckian $f$ depends on the precise definition of what a minimally charged object is allowed to be. In this context, it may be crucial to understand the gravitational dynamics of the composite effective string introduced above.

\section*{Acknowledgments}

We thank Patrick Mangat, Eran Palti, Fabrizio Rompineve, Pablo Soler, and Alexander Westphal for helpful discussions. This work was supported by the DFG Transregional Collaborative Research Centre TRR 33 ``The Dark Universe'' and the HGSFP. LW is partially supported by the Advanced ERC grant SM-grav, No 669288.

\bibliography{AxionWGCbib}
\bibliographystyle{JHEP}
\end{document}